\documentclass[reprint, superscriptaddress, amsmath, amssymb, aps,pra]{revtex4-2}

\usepackage{graphicx}
\usepackage{dcolumn}
\usepackage{bbm}
\usepackage{hyperref}

\usepackage{tikz}
\usepackage{lipsum}
\usepackage{todonotes}
\usepackage{braket}
\usepackage{bbm}

\newcommand{\e}{\mathrm{e}}
\renewcommand{\i}{\mathrm{i}}
\newcommand{\eqdef}{\overset{\text{def.}}{=}}

\newcommand{\id}{\hat{\mathbbm{1}}}
\newcommand{\tr}{\mathrm{tr}}
\renewcommand{\d}{\mathrm{d}}

\newcommand{\densmat}{\hat{\rho}}
\newcommand{\h}{\hat{H}}

\newcommand{\hint}{\hat{H}_{\text{int}}}

\renewcommand{\a}{\hat{a}}
\newcommand{\ad}{\hat{a}^{\dagger}}
\newcommand{\ak}{\hat{a}_k}
\newcommand{\akd}{\hat{a}_k^{\dagger}}

\newcommand{\hc}{\text{h.c.}}

\renewcommand{\Re}{\mathrm{Re}}

\begin{document}

\title{Connectivity matters\\ Impact of bath modes ordering and geometry in open quantum system simulation with Tensor Network States}

\author{Thibaut Lacroix}
\email{thibaut.lacroix@uni-ulm.de}
\affiliation{
Institut für Theoretische Physik \& IQST, Albert-Einstein-Allee 11, Universität Ulm, D-89081 Ulm, Germany
}

\author{Brendon W. Lovett}
\affiliation{%
 SUPA, School of Physics and Astronomy, University of St Andrews, St Andrews KY16 9SS, UK}

\author{Alex W. Chin}
 \affiliation{
 Sorbonne Universit\'{e}, CNRS, Institut des NanoSciences de Paris, 4 place Jussieu, 75005 Paris, France
}%

\begin{abstract}
    Being able to study the dynamics of quantum systems interacting with several environments is important in many settings ranging from quantum chemistry to quantum thermodynamics, through out-of-equilibrium systems. 
    For such problems tensor network-based methods are state-of-the-art approaches for performing numerically exact simulations. 
    However, to be used efficiently in this multi-environment and non-perturbative context, these methods require an optimized choice of the topology of the wave-function Ans\"atze. This is often done by analysing cross-correlations between different system and environment degrees of freedom.
    Here, we show for canonical model Hamiltonians that simple orderings of bosonic environmental modes, which enable the joint \{System + Environments\} state to be written as a matrix product state, considerably reduce the bond dimension required for convergence despite introducing long-ranged interactions.
    These results suggest that complex correlation analyses for tweaking tensor networks topology (e.g. entanglement renormalization) are usually not necessary and that tree tensor network states are sub-optimal compared to simple matrix product states in many important models of physical open systems.
\end{abstract}

\maketitle

\section{Introduction}
The theory of open quantum systems describes how all real-world quantum systems are influenced by their interactions with external, often uncontrollable, degrees of freedom~\cite{breuer_theory_2007, weiss_quantum_2012, rivas_open_2011}.
These external degrees of freedom that influence the open systems' dynamics can be of a different nature (bosonic, fermionic, spins), have a different character (vibrational, electromagnetic) or temperature and are thus said to belong to separate `environments'.
Various methods have been developed to describe the dynamics of the system of interest's populations and coherences.
In cases where the inner dynamics of the environment can be neglected, Lindblad and Redfield master equations are tools of choice~\cite{haroche_exploring_2006, rivas_open_2011}. 
However, when this inner dynamics cannot be neglected, more involved methods are needed~\cite{tanimura_numerically_2020, lambert_modelling_2019, makri_tensor_1995, strathearn_efficient_2018, cygorek_simulation_2022, tamascelli_efficient_2019, somoza_dissipation-assisted_2019, meyer_studying_2012}.
A number of these methods are numerically exact and rely on tensor networks~\cite{orus_practical_2014, shi_classical_2006} to efficiently encode this environmental dynamics, for example in the form of an influence functional~\cite{makri_tensor_1995, strathearn_efficient_2018, cygorek_simulation_2022, link_open_2024} or the total state of the \{System + Environment\}~\cite{tamascelli_efficient_2019, meyer_studying_2012, somoza_dissipation-assisted_2019}.
Indeed, tensor networks give a compressed and computationally efficient representation of these objects.

An example, of particular interest for quantum technologies, comes from quantum thermodynamics where one might be interested in studying the heat exchange of a quantum system interacting with hot and cold reservoirs (e.g. a quantum heat engine)~\cite{deffner_quantum_2019}.
Another example comes from molecular systems where electronic excitations can effectively couple to several different vibrational environments~\cite{schroder_tensor_2019, dunnett_influence_2021} or environments of a different nature~\cite{le_de_extending_2024}.

Having several environments interacting with the system leads to an increase of the required computational resources needed to perform an accurate simulation, because the degree of correlation between the system and the environmental degrees of freedom grows.
For instance, for Matrix Product States (MPS) the computational cost associated with the computation of observables is related to their bond dimension, which is itself related to the degree of correlation (or the amount of entanglement) between bi-partitions of the many-body system-environment wave function.
For methods based on tensor network states (TNS), a few techniques exist to reduce the computational cost (the bond dimension) of such simulations.
They are either based on -- what we coin -- `structural renormalisation', i.e. changing the structure of the TNS Ansatz to locally lower the entanglement~\cite{schroder_tensor_2019,  mendive-tapia_optimal_2023, hikihara_automatic_2023}, or on `clustering' of environmental modes, i.e. changing the order of modes for a fixed geometry~\cite{rams2020breaking, li_fly_2022, kohn_efficient_2021, li_optimal_2024}.
The structural renormalisation approach relies on expressing the joint quantum state as a tree tensor network (TTN) state and then analysing the entanglement properties of system-environment and environment-environment partitions.
From this analysis, a new topology of the TTN is deduced~\cite{mendive-tapia_optimal_2023, hikihara_automatic_2023} or a new layer of so-called entanglement renormalising (ER) tensors is constructed and placed in between the system and the environments to lower the entanglement but keep correlations between environments~\cite{schroder_tensor_2019}.
This approach necessitates a thorough analysis of the entanglement structure between the system and the environments and between the environments themselves before being able to write the TTN state.
On the other hand, the methods relying solely on the reordering of the environmental modes preserve a simple TTN or MPS structure.
How environmental modes are ordered in the MPS representation of the quantum state is \emph{a priori} arbitrary.
One just has to make sure that the operators in the Matrix Product Operators (MPOs) applied to the state follow the same ordering; this might make their form more complex.
Nevertheless, it does not mean that all orderings of the environmental modes are equivalent.
In the case of bosonic modes, `On the Fly Swapping' is a procedure applied during the sweeps of the (real or imaginary) time evolution where neighbouring tensor sites are swapped if it results in decreasing a given cost function (the entanglement entropy or the truncation error)~\cite{li_fly_2022}.
Even though this method is a practical way of improving computational performance, it needs to be applied at each time step of the state evolution.
In the case of fermionic bath modes, the reordering is particularly simple and consists of alternating bath modes corresponding to filled orbitals with modes corresponding to empty ones~\cite{kohn_efficient_2021, kohn_quench_2022}.
Whereas the reordering procedure for bosonic degrees of freedom is a dynamic procedure, this fermionic reordering is static and done once and for all when writing down the initial state.

A natural question then emerges: is there a similar efficiency-increasing static reordering of bath modes in the bosonic case?
If this were the case, it would spare complex beforehand-analyses of system-environments correlations or dynamical updates of the ordering of environmental modes.
In this paper, we focus on methods where a TNS describes the whole \{System + Environment(s)\} wave-function.
We use a specific instance of such methods called \emph{Thermalized-Time Evolving Density operator with Orthogonal Polynomials Algorithm} (T-TEDOPA)~\cite{chin_exact_2010, prior_efficient_2010, woods_simulating_2015, tamascelli_efficient_2019, lacroix_unveiling_2021}.
We study here the influence of different bosonic mode arrangements on the efficiency of open quantum system simulations.
We initially conjecture that entanglement in MPS states can be lowered by diminishing the \emph{`correlation length'} between two environmental excitations created at the same time.
Investigating the convergence of different arrangements of bath modes we are able to give a negative answer to this hypothesis.
This exploration brings us to study another property of tensor network states that we show to be related to entanglement: the connectivity of the system with its environments.
In the case of a single environment, it has been argued based on empirical results that TTN states encode many-body correlations more efficiently than MPS~\cite{dorfner_comparison_2024}.
In this paper, considering the case of a system coupled to several environments, we bring evidence that adopting a TNS geometry with a single \emph{`interface'} between the system and the environments to which it connects reduces the amount of entanglement in the state.
Contrary to previous comparative studies between MPS and TTN, a strength of this paper is the comparison of different MPS and TTN representations of a quantum state on an equal footing.
Indeed, the same numerical method (T-TEDOPA) and the same time-evolution algorithm (the Time-Dependent Variational Principle) are used for all TNS.

The paper is organized as follow: The T-TEDOPA numerical method is presented in Section~\ref{sec:TEDOPA}.
In Section~\ref{sec:2baths} we define different orderings of the environmental modes and compare their efficiency for a system coupled to two environments.
The case of three environments is studied in Section~\ref{sec:3baths}.
All these results are discussed in Section~\ref{sec:discussion}.

\section{Numerical method\label{sec:TEDOPA}}
When considering a quantum system linearly coupled to a continuum of bosonic modes (e.g. the electromagnetic field or vibrational modes in molecules), we need to discretize the environment in order to perform numerical simulations of the \{System + Environment\} dynamics.
Instead of sampling the infinitely many normal modes of the environment to keep only a discrete set of modes, a chain mapping approach can be used to easily keep all relevant bath modes and at the same time generate a discrete representation of the environment~\cite{chin_exact_2010}.
This method consists of using an exact unitary transformation defined through a family of orthonormal polynomials that transforms a continuous bosonic environment into a semi-infinite chain and is known as Time Evolving Density operator with Orthonormal Polynomials Algorithm (TEDOPA)~\cite{prior_efficient_2010, woods_simulating_2015}.
At zero temperature, this chain-mapped environment is well-suited for a representation of the joint \{System + Bath\} wave-function as a MPS because the bath is now made of discrete modes, and all the couplings of the joint system are local.
Moreover, in this representation, a MPS reflects the underlying structure of the Hamiltonian.
This method is very efficient for representing the OQS non-perturbatively and in the non-Markovian regime, allowing the evolution of the full wave-function of the system and its environment to be simulated.

We consider a bosonic bath described by the Hamiltonian $\h_B$ and interacting with a reduced system of interest with the Hamiltonian $\hint$
\begin{align}
    \h_B + \hint = &\int_{0}^{\omega_c} \d \omega \omega \ad_\omega\a_\omega + \hat{A}\int_{0}^{\omega_c} \d \omega \sqrt{J(\omega)}(\hat{a}_\omega + \hat{a}_\omega^\dagger)\ ,
\end{align}
where $\ad_\omega$ creates a bosonic excitation at the frequency $\omega$, $\omega_c$ is the bath cut-off frequency, $J(\omega)$ is the bath spectral density (SD), and $\hat{A}$ is a system operator.

We can introduce the following transformation of the bath operators
\begin{equation}
    \hat{a}_\omega = \sum_{n=0}^{\infty} U_n(\omega)\hat{b}_n
    \label{eq:def_chainmodes}
\end{equation}
where $U_n(\omega)$ is defined with orthonormal polynomials $P_n$
\begin{equation}
    U_n(\omega) = \sqrt{J(\omega)} P_n(\omega) \ .
    \label{eq:unitary-polynomials}
\end{equation}
Equation~(\ref{eq:def_chainmodes}) expresses the decomposition of the continuum of independent bosonic modes onto a new infinite set of discrete modes.
The unitarity of the transformation imposes an orthogonality relation for the polynomials
\begin{align}
    \int_0^{\omega_c} \d \omega U_n(\omega) U_m(\omega) &= \int_0^{\omega_c} \d \omega J(\omega) P_n(\omega)P_m(\omega) = \delta_{n,m}\ .
    \label{eq:orthogonality}
\end{align}
This orthogonality relation defines the family of polynomials used for the transformation.
Thus, the chosen polynomials depend on the bath spectral density $J(\omega)$.

Another useful property of these polynomials is that they obey a recurrence relation
\begin{align}
    P_n(\omega) &= (\omega - A_{n-1})P_{n-1}(\omega) + B_{n-1}P_{n-2}(\omega)\ ,
    \label{eq:recurrence}
\end{align}
where $A_n$ is related to the first moment of $P_n$ and $B_n$ to the norms of $P_n$ and $P_{n-1}$~\cite{chin_exact_2010}.
This recurrence relation can be used to construct the polynomials with the conditions that $P_0(\omega) = 1$ and $P_{-1}(\omega) = 0$.

If we apply this unitary transformation to the interaction Hamiltonian
\begin{align}
\hat{H}_\text{int} &= \sum_n\hat{A} \int_0^{\omega_c} \d \omega J(\omega)P_n(\omega)(\hat{b}_n + \hat{b}_n^\dagger)\nonumber\\
&= \hat{A} (\hat{b}_0 +\hat{b}_0^\dagger)\ ,
\label{eq:short-range}
\end{align}
we notice that the system only couples to the first mode of the transformed bath.

The same transformation applied to the bath Hamiltonian yields to the following nearest-neighbour hopping Hamiltonian where $\omega_n$ is the energy of chain mode $n$ and $t_n$ is the coupling between mode $n$ and $n+1$
\begin{equation}
    \hat{H}_B = \sum_n \omega_n \hat{b}_n^\dagger\hat{b}_n + t_n (\hat{b}_n^\dagger\hat{b}_{n+1} + \hat{b}_{n+1}^\dagger\hat{b}_{n}) \ .
    \label{eq:tightbinding}
\end{equation}

\begin{figure}
    \centering
    \includegraphics[width=0.7\columnwidth]{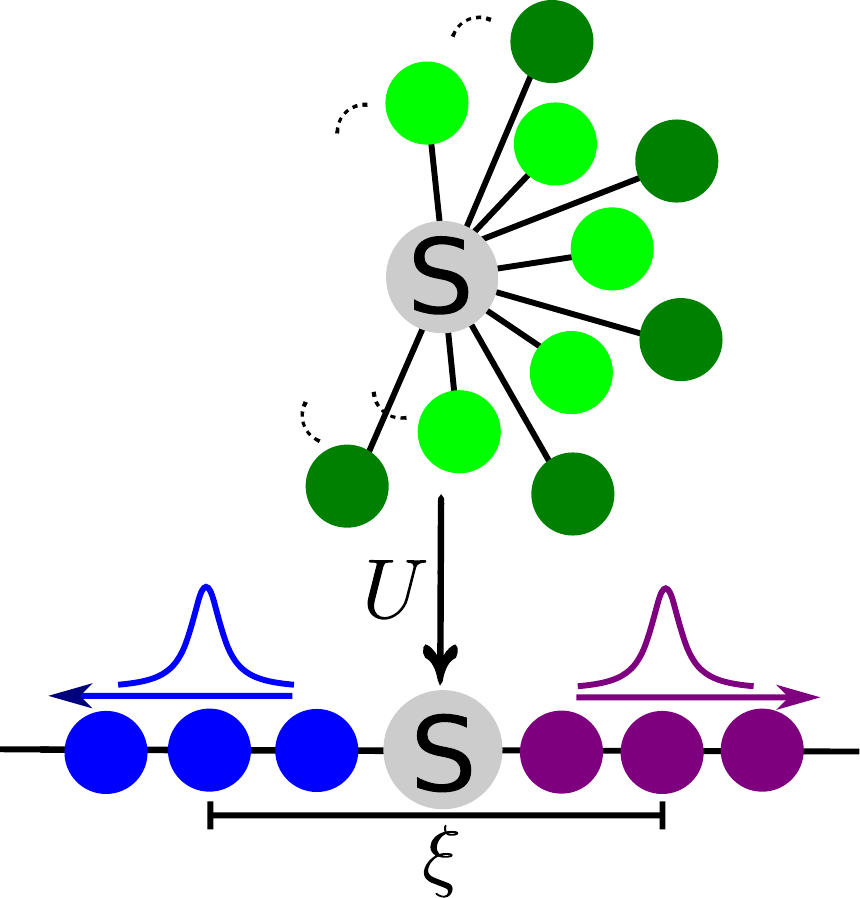}
    \caption{Schematic drawing of the unitary transformation $U$ -- the chain mapping -- changing two baths of independent bosonic modes into two chains with nearest neighbour interactions. The correlation length $\xi(t)$ measures the distance in between two dynamically correlated environmental excitations.}
    \label{fig:chainmapping}
\end{figure}

From the new bath and interaction Hamiltonians of Eqs.~(\ref{eq:short-range}) and (\ref{eq:tightbinding}) we can see that the unitary transformation $U_n(\omega)$ transforms the bosonic environment composed of a continuum of independent modes --- which is called the star geometry of the environment --- into a semi-infinite chain of interacting modes.
This one-dimensional nearest-neighbour Hamiltonian is well suited to a representation in the form of a MPO or a TTN.
Figure~\ref{fig:chainmapping} shows such a mapping in the case of two environments.
The system is coupled directly to all the normal bath modes, but after this transformation it is only coupled to the first chain mode $n=0$.
An excitation injected into this mode, i.e. the system dissipating energy into the environment, can then travel along the chain as a wavefront~\cite{tamascelli_excitation_2020}.
For practical purposes, the chain is truncated to contain only $N_m$ modes.
The only constraint is that this number should be large enough that the wavefront generated by the interaction with the system does not reflect back to the beginning of the chain.
Retaining only a finite number of modes in the chain representation also corresponds to considering a finite set of normal modes in the original bath.
Truncating the chain corresponds to an optimal sampling procedure where the bath modes are not sampled arbitrarily~\cite{de_vega_how_2015}.
It has recently been shown that the procedure can be extended to describe finite temperature baths by replacing the finite temperature bath by a fictitious zero temperature one with a temperature-dependent coupling to the system and a spectrum allowing negative energies~\cite{tamascelli_efficient_2019,dunnett_simulating_2021,riva2023thermal}.
This `trick' enables the use of MPS to describe the quantum states of systems coupled to finite temperature Gaussian baths and is the cornerstone of the T-TEDOPA method.

We carry out simulations mainly using a one-site implementation of the \emph{Time-Dependent Variational Principle} (1TDVP)~\cite{paeckel, dunnett_angusdunnettmpsdynamics_2021}.
This implementation of the TDVP has the advantages of conserving the unitarity of the evolution, the total energy, and has a better scaling of its computational cost with respect to the local Hilbert space dimension of the bath modes than its multi-site alternative.
However, it requires the MPS to be embedded in a manifold of fixed bond dimension $D$.
To perform TDVP simulations, we express the Hamiltonian describing the system, its environment(s) and their interaction as a MPO.
All numerical results presented here were produced using our open source Julia software package \texttt{MPSDynamics.jl}~\cite{dunnett_angusdunnettmpsdynamics_2021, mpsdynamicsjl_2024}, where TDVP for MPS and TTN is implemented, as well as methods to generate chain-mapping coefficients.
The package is freely available at \href{https://github.com/shareloqs/MPSDynamics.jl}{https://github.com/shareloqs/MPSDynamics.jl}.

The analysis performed in this paper is solely possible with methods working with the full wave function of the system and the environment.

\section{Two environments: Mode arrangements\label{sec:2baths}}
In several cases the quantum system of interest is interacting with two baths.
A first instance comes from the study of thermodynamic properties and dynamics of quantum systems. 
For example, this occurs when studying heat flows when a quantum system is between a cold bath and a hot bath or when the system is thermalizing~\cite{dunnett_efficient_2021, fux_thermalization_2022}.
This category of situations is important in the context of quantum thermal machines.
Another example comes when describing propagating and counter-propagating modes in 1D for spatially extended systems~\cite{lacroix_unveiling_2021}.
More generally, the two bosonic baths could be of different nature, for example one representing a vibrational environment and the other an electromagnetic environment~\cite{gribben_exact_2022, le_de_extending_2024}.
There, the two environments could play a different role, and their effects can simply add up; alternatively, new features that do not result from such simple addition can emerge. Such behaviour is called non-additive\cite{wertnik2018optimizing}, and has been observed in several studies of ultrafast molecular photophysics \cite{schroder_tensor_2019,dunnett_influence_2021}.  
In any case, the problem of a quantum system interacting with two bosonic environments is more than just a case study and has physical relevance.

When the two environments have similar characteristic timescales, energy dissipation from the system results in the generation of environmental excitations that are then propagating in their respective chains in a dynamically correlated way~\cite{calabrese_evolution_2005}.
It thus appears `natural' to define a \emph{`correlation length'} $\xi$ between these excitations, namely the number of MPS's sites in between them.
The hypothesis that directly follows is that the degree of entanglement (described by the bond dimension) grows with this correlation length.
Though not formulated in this way, this hypothesis forms a main conclusion of the work previously performed on mode reordering in the context of fermionic baths~\cite{kohn_quench_2022}. Indeed, several promising numerical techniques for simulating open system dynamics with MPS or MPO techniques have been developed to manage, or remove, these numerically costly correlations through the addition of dissipation or dephasing to the environment \cite{somoza_dissipation-assisted_2019, wojtowicz_open-system_2020}.

Hence, in this section we introduce three different orderings of bath modes (depicted in Fig.~\ref{fig:arrangements}) that display three different behaviours for the correlation length $\xi$, and we compare the minimum bond dimension they require in order to reach convergence in 1TDVP simulations.

\subsection{Left-Right arrangement}
\begin{figure}
    \centering
    \includegraphics[width=\columnwidth]{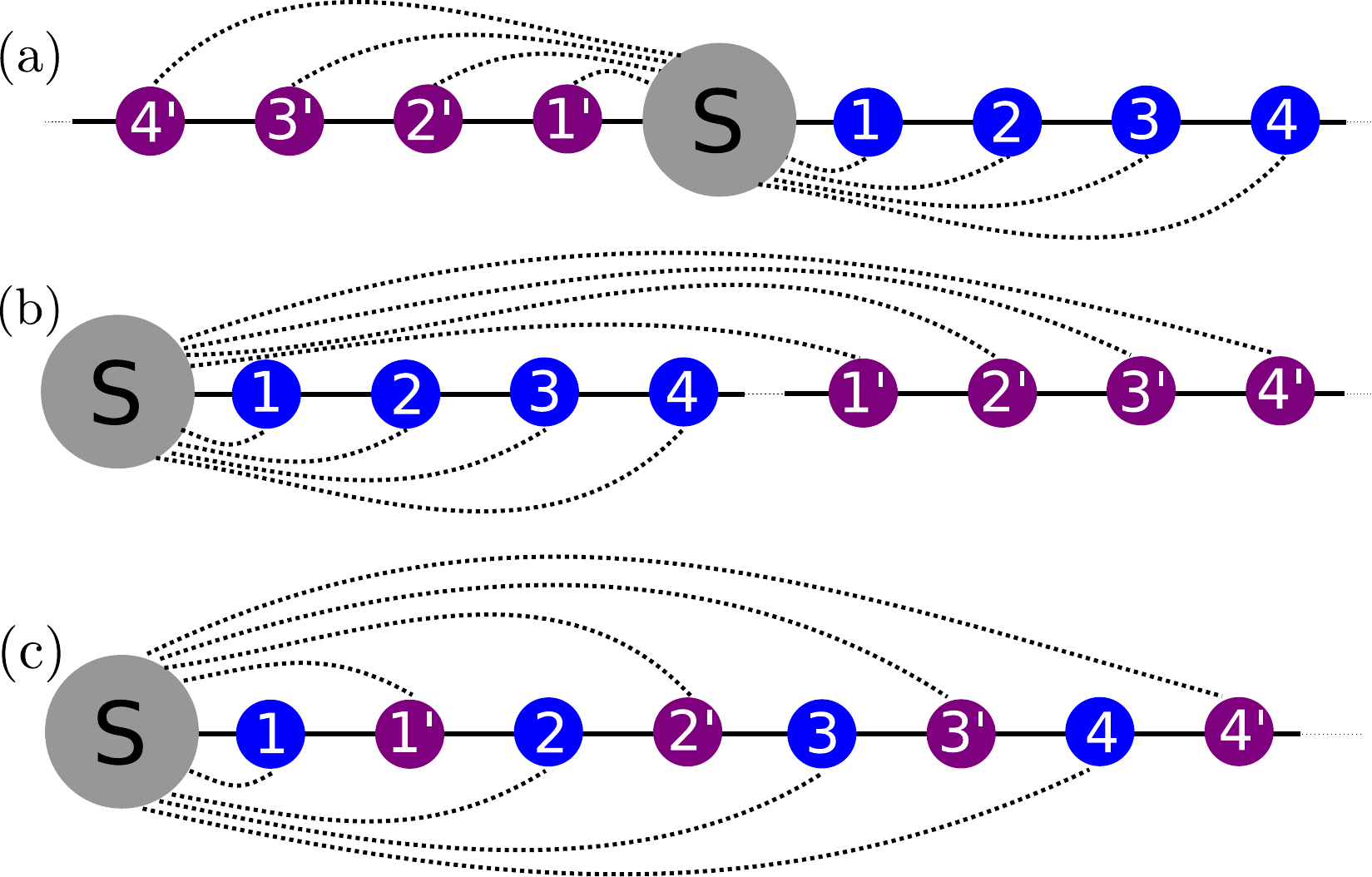}
    \caption{Diagram of the different mode arrangements in the MPS for a system interacting with two bosonic baths. The dotted lines represent the long-range couplings between the system and chain modes. (a) \emph{`Left-Right'} arrangement of the chain-mapped environment. Chain modes are symmetrically placed on both sides of the system. The correlation length between the excitations of the two chains grows linearly in time. (b) \emph{`Successive'} arrangement of the chain-mapped environment. One chain connects directly to the system and the second one is appended to the first one. The correlation length between the excitations of the two chains is a constant of the order of the length of the chain. (c) \emph{`Interleaved'} arrangement of the chain-mapped environment. Chain modes are alternating and thus coupling to their next nearest neighbours. The correlation length is constant of the order of 1.}
    \label{fig:arrangements}
\end{figure}

Usually, the MPS and MPO representation of the joint \{System + Environments\} state and Hamiltonian follows what we call a \emph{`Left-Right'} ordering where the system is placed in between the two chains (see Fig.~\ref{fig:arrangements}).
This arrangement is `intuitive' because it reproduces the structure of the Hamiltonian and, thus, often used when describing a quantum system interacting with two independent baths ~\cite{dunnett2021matrix, kohn_efficient_2021}.
However, this ordering of the environmental modes might lead to a non-optimal maximal bond dimension.
Indeed, the initial environmental excitations created by the environment propagate towards the end of their respective chains but are dynamically correlated.
Hence, as time passes, these excitations that are highly correlated move further apart thus \emph{a priori} requiring a higher bond dimension for the MPS as the correlation length grows linearly with time $\xi \propto t$.
Figure~\ref{fig:chainmapping} shows a representation of this correlation length.
Notice that the Left-Right MPS is `isomorphic' to a TTN state with two branches: one for each environment.

\subsection{Successive arrangement}
An alternative ordering of the chain modes corresponds to placing the system at one end of a chain and the other chain at the other end (i.e. concatenate the two chains), as shown in Fig.~\ref{fig:arrangements}~(b).
We coin this ordering of the bath modes \emph{`Successive'}.
This might seem counter-intuitive as the initial bath excitations thus immediately induce a correlation over a chain-long region.
However, the correlation length stays approximately constant $\xi \approx N_m$ as the excitations propagate along the chain.

\subsection{Interleaved arrangement}
To reduce the bond dimension a solution would be to bring closer the modes that are highly correlated with one another.
In the Successive arrangement case, compared to the Left-Right one, this is done by having a fixed distance of $\xi\approx N_m$ between the correlated excitations.
An improvement could be achieved by interleaving the two chains together into a single one.
This is done by alternating the mode of one chain with the corresponding mode of the other one, as shown in Fig.~\ref{fig:arrangements}~(c).
Thus, correlated excitations are now separated by a fixed distance of $\xi\approx 1$ thus diminishing the correlation length.
This type of mode ordering has already been applied to fermionic problems, namely the Anderson impurity model, where one bath describes empty orbitals and the other one filled orbitals.

\subsection{Comparison}\label{sec:2bathsComparison}

To compare these three different mode arrangements, we are first going to apply them to a well known and studied OQS model, the Independent Boson Model (IBM), and later on to the Spin Boson Model (SBM).
The Spin Boson Model is a paradigmatic model of OQS where a single two-level system (TLS) interacts linearly with a bosonic environment
\begin{align}
    \h =& \left(\frac{\epsilon}{2}\hat{\sigma}_z + \frac{\Delta}{2}\hat{\sigma}_x\right)\otimes\id_{E} + \id_{S}\otimes\int_0^{+\infty}\omega\ad_\omega\a_\omega \d\omega \nonumber\\
    &+ \frac{\hat{\sigma}_x}{2}\otimes\int_0^{+\infty} \left(g_\omega\a_\omega + \hc\right)\d\omega\ ,\label{eq:SBM}
\end{align}
where $\hat{\sigma}_i$ are the Pauli matrices, $\ad_\omega$ ($\a_\omega$) is a bosonic creation (annihilation) operator for the mode of frequency $\omega$, $\id_S$ ($\id_E)$ is the identity operator on the Hilbert space of the system (environment), $\epsilon$ is the TLS energy gap and $\Delta$ its tunnelling term, and $g_\omega$ is the coupling strength between the TLS and the mode of frequency $\omega$.
When the energy gap is zero ($\epsilon = 0$) the SBM reduces to the so-called Independent Boson Model where the system and interaction Hamiltonians commute.
Contrary to the SBM, the IBM is analytically solvable.
In the IBM, because $\left[\h_S, \hint \right] = 0$, the system's populations are invariant. 
The system only experiences pure dephasing of its initial coherences and is not subjected to decay.

\subsubsection{Independent Boson Model}
\begin{figure*}
    \centering
    \includegraphics[width=\textwidth]{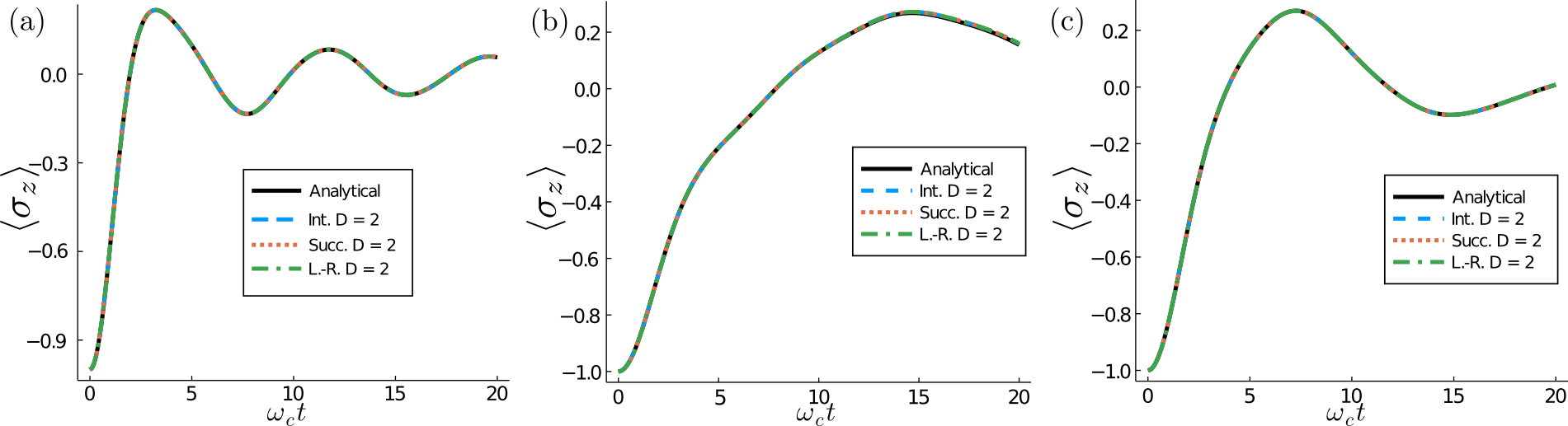}
    \caption{Dynamics of the $\langle \sigma_z\rangle$ for the Interleaved (Int.), Successive (Succ.) and Left-Right (L.-R.) chain arrangements for a bond dimension $D = 2$ for the Independent Boson Model. (a) The model parameters are $\Delta = 0.8\omega_c$, $\beta = \infty$, $\alpha = 0.2$. (b) The model parameters are $\Delta = 0.2\omega_c$, $\beta = \infty$, $\alpha = 0.1$. (c) The model parameters are $\Delta = 0.4\omega_c$, $\omega_c\beta = 10$, $\alpha = 0.1$. The three different arrangements are able to recover the analytical results at zero and finite temperature for the first non trivial bond dimension.}
    \label{fig:IBM}
\end{figure*}
We first show that all arrangements are able to accurately describe system's dynamics at zero and finite temperature by considering the IBM.
The Hamiltonian we consider has two bosonic baths
\begin{align}
    \h &= \frac{\Delta}{2}\hat{\sigma}_x + \sum_{i=1}^{2}\int_{\mathbb{R}} \omega_k \ak^{i \dagger}\ak^i \d k + \frac{\hat{\sigma}_x}{2}\sum_{i=1}^{2}\int_{\mathbb{R}}\left(g_k\ak^i + \hc\right)\d k\ ,
\end{align}
where $\ak^{i \dagger}$ creates an excitation in bath $i$ with energy $\omega_k$, and $g_k$ is the coupling strength between bath mode $k$ and the system for both baths.
Because the energies of the modes $\omega_k$ and the coupling constants $g_k$ are the same for both baths, this model could also be written as a single bath IBM with a coupling twice as strong.
The expectation value $\langle\sigma_z\rangle$ is given by~\cite{breuer_theory_2007, weiss_quantum_2012}
\begin{widetext}
\begin{align}
    \langle\sigma_z(t)\rangle &= \cos(\Delta t)\Re\left[ \rho_{\uparrow\downarrow}(0) \exp\left(-\int_{0}^{\infty} \frac{J(\omega)}{\omega^2}\left(1 - \cos(\omega t)\right)\coth\left(\frac{\beta\omega}{2}\right) \right)\right]\ ,
\end{align}
\end{widetext}
where $\rho_{\uparrow\downarrow}(0)$ is the initial coherence of the system, $J(\omega) \eqdef \sum_k|g_k|^2\delta(\omega - \omega_k) = 2\alpha\omega H(\omega_c - \omega)$ is chosen to be an Ohmic SD and $\beta = (k_B T)^{-1}$ is the bath inverse temperature.
Figures~\ref{fig:IBM}~(a) and \ref{fig:IBM}~(b) show that the Left-Right, Successive and Interleaved arrangements all recover the analytical expression for the smallest non-trivial bond dimension $D = 2$ for widely different values for $\Delta$ and $\alpha$.
This small bond dimension is expected as the analytical solution of state of the system (for the initial state given below) is known to be an entangled state between the system and a displaced environment 
\begin{align}
    \ket{\psi(t)} = &\frac{\e^{-\i\theta(t)}}{\sqrt{2}}\bigg( \ket{\uparrow_x}\bigotimes_{i=1}^{2} \hat{D}(\{\alpha_k^i(t)\})\ket{\{0\}_{k,i}}\nonumber\\
    &- \ket{\downarrow_x}\bigotimes_{i=1}^2\hat{D}(\{-\alpha_k^i(t)\})\ket{\{0\}_{k,i}}\bigg)
\end{align}
where $\theta(t)$ and $\alpha_k^i(t) = \alpha_k(t)$ depend solely on the SD $J(\omega)$ and the energies of bath modes $\omega_k$ and $\hat{D}(\{\alpha_k\})$ is a multi-mode displacement operator.
Given that the environment state is simply a displaced state, manifesting no entanglement of its own, the information about the joint state can be stored with a single bit of information, hence the value of the bond dimension $D = 2$.
Similarly, Fig.~\ref{fig:IBM}~(c) shows that the analytical behaviour is also recovered at finite temperature for the first non-trivial bond dimension.
The initial joint state is a product state 
\begin{align}
    &\Biggl \{
    \begin{matrix}
    &\ket{\downarrow_z}\otimes\ket{\{0\}_k}& \text{if} & \beta = \infty\\
    &\ket{\downarrow_z}\bra{\downarrow_z}\otimes\exp\left(-\beta\h_B\right)/Z & \text{if} & \beta \neq \infty
    \end{matrix}\ ,
\end{align}
where $\ket{\downarrow_z}$ is the eigenstate of $\hat{\sigma}_z$ associated with the eigenvalue $-1$, and $Z$ is the bath partition function.

\subsubsection{Correlated Environment}
Now that we have established that our three different arrangements are able to accurately describe OQS dynamics, we are going to study the convergence behaviour of these different orderings with a non-trivial model: a system interacting with propagating modes.
We consider a 1D system composed of two sites labelled by $\gamma$ placed in space at the position $r_\gamma$.
These two sites are interacting with a common bosonic bath of plane waves labelled by the wave-vector $k \in \mathbb{R}$.
The corresponding Hamiltonian is 
\begin{align}
    \h &= \sum_{\gamma}E_\gamma \hat{f}_\gamma^\dagger\hat{f}_\gamma + \omega_0\left(\hat{f}_\gamma\hat{f}_{\gamma+1}^\dagger +\hc \right) + \int_{\mathbb{R}} \omega_k \akd\ak \d k \nonumber\\
    &+ \sum_{\gamma}\hat{f}_\gamma^\dagger\hat{f}_\gamma\int_{\mathbb{R}}\left(g_k\e^{\i k r_\gamma}\ak + \hc\right)\d k\ ,
\end{align}
where $\hat{f}_\gamma^\dagger$ creates an excitation on site $\gamma$, $\akd$ creates a plane wave of energy $\omega_k = |k|c$ with $c$ the phonon speed (and $\hbar = 1$), and $g_k\e^{\i k r_\gamma}$ is a coupling strength between bath mode $k$ and the site $\gamma$.
In that case, the bosonic environment is mapped to two chains, one for the propagating bath modes and the other one for the counter-propagating modes, with long-ranged couplings with the system~\cite{lacroix_unveiling_2021, dunnett_influence_2021}.
In terms of the MPO representation of the Hamiltonian, the long-ranged couplings lead to an increase of the operator bond dimension which has been proven to scale with the system size only~\cite{lacroix_unveiling_2021}.
We comment on the fact that this is a particular instance of a system coupled to two bosonic baths (they happen to be described by the same parameters up to complex conjugation).
We consider this specific type of environment because it is general and can be easily extended to the case of two bosonic baths with different parameters.
Furthermore, in that case, the dynamics of the two environments is highly correlated.
The results shown hereafter are for degenerate sites $E_\gamma = 0$ at positions $k_c r_1 = 0$ and $k_c r_2 = 5$ for different values of coupling strengths to the bath $\alpha$ (the SD is again taken to be Ohmic) and coherent coupling $\omega_0$.
Initially the system state is localised on the first site
\begin{align}
    &\Biggl \{
    \begin{matrix}
    &\ket{r_1}\otimes\ket{\{0\}_k}& \text{if} & \beta = \infty\\
    &\ket{r_1}\bra{r_1}\otimes\exp\left(-\beta\h_B\right)/Z & \text{if} & \beta \neq \infty
    \end{matrix}\ ,
\end{align}
where $\ket{r_1}$ is the state corresponding to site 1 being occupied, i.e. $\bra{r_1}\hat{f}_\gamma^\dagger\hat{f}_\gamma\ket{r_1} = \delta_{\gamma, 1}$.
\begin{figure*}
    \centering
    \includegraphics[width=\textwidth]{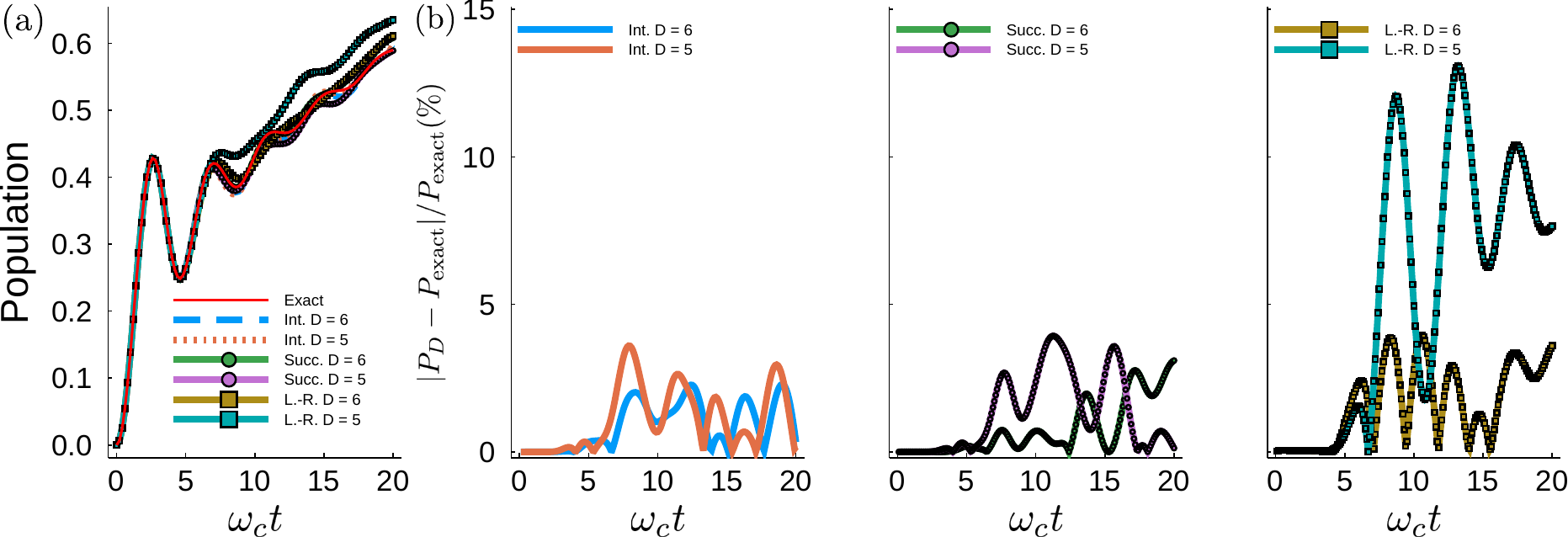}
    \caption{ (a) Population of the second site in the correlated environment model. Int., Succ. and L.-R. respectively stand for an Interleaved, a Successive, and a Left-Right chain arrangement. All the arrangements are converged for $D = 15$ which is labeled as Exact. (b) Relative error with respect to the exact $D = 15$ case. Generally the Left-Right arrangement converges for higher bond dimensions than the Interleaved and Successive arrangements that have a similar convergence behaviour. The model parameters are $E_\gamma = 0$, $\alpha = 0.2$, $\omega_0 = 0.4\omega_c$, $\beta = \infty$.}
    \label{fig:comparison}
\end{figure*}

Figure~\ref{fig:comparison} shows the population of the second site at zero temperature for a strong system-bath coupling $\alpha = 0.2$ and a strong tunneling energy $\omega_0 = 0.4\omega_c$, obtained with the three arrangements for different bond dimensions.
The different mode arrangements are all converged for $D = 15$ but only the Interleaved one is represented at this bond dimension for readability.
We can see that Left-Right is generally performing worse than the Successive and Interleaved arrangements.
For non-converged values of $D$, this arrangement exhibits dynamics that are further away from the converged one than the Successive and Interleaved arrangements.
These two arrangements look similar but Interleaved seems to catch the coherent oscillations better for $D=6$.
The inset of Fig.~\ref{fig:comparison} shows the relative error between the converged population and the population computed with lower bond dimensions for the three arrangements.
From these examples, we can rule out the explanation of better convergence depending on shorter correlation length because the Interleaved and Successive arrangements have very similar convergence properties despite having an order of magnitude difference in their correlation lengths.
Nevertheless, these two orderings are still converging faster than the usual Left-Right one.
This result might be surprising as these arrangements do not reproduce the structure of the underlying Hamiltonian.

\begin{figure}
    \centering
    \includegraphics[width=\columnwidth]{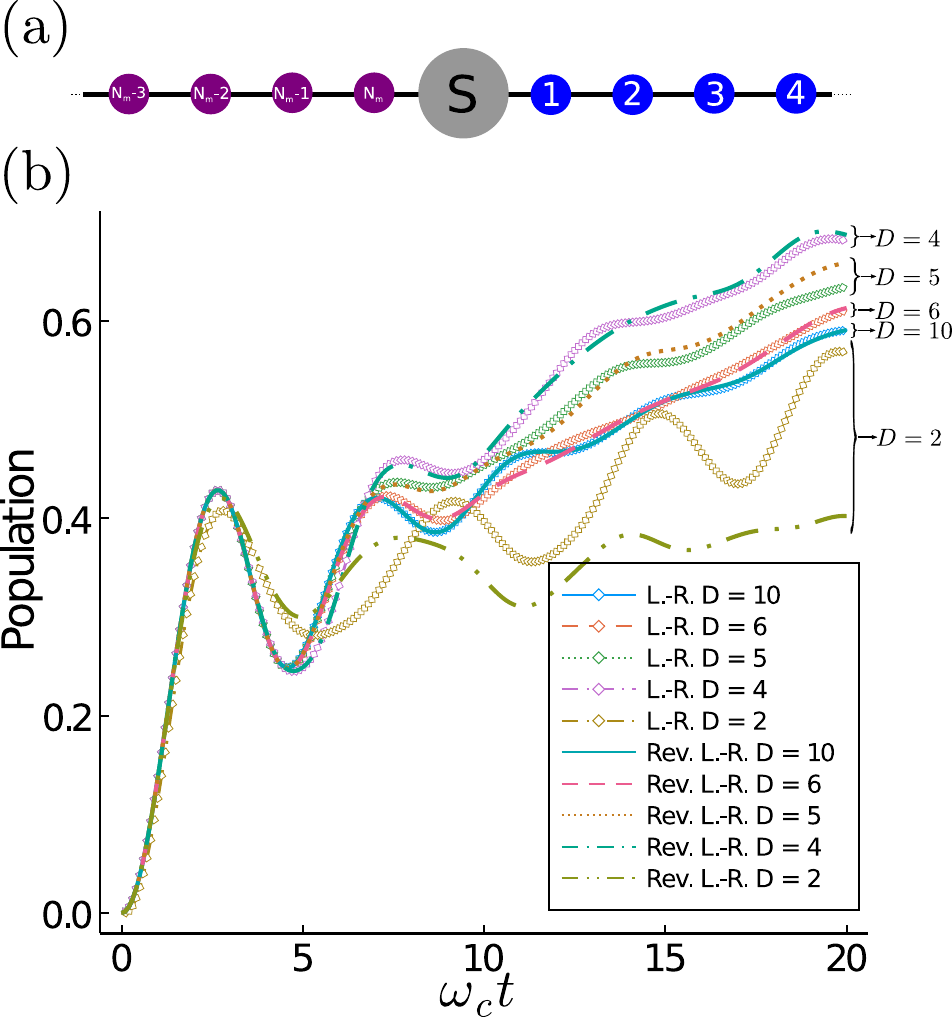}
    \caption{(a) \emph{`Reverse Left-Right'} arrangement of the chain-mapped environment. Chains modes are symmetrically placed on both side of the system but on one chain they are in the reverse order. The correlation length between the excitations of the two chains remains constant and is the same as in the `Successive' ordering. (b) Comparison of the Left-Right (L.-R.) and Reverse Left-Right (Rev. L.-R.) ordering of chains modes for the same parameters as in Fig.~\ref{fig:comparison} ($E_\gamma = 0$, $\alpha = 0.2$, $\omega_0 = 0.4\omega_c$, $\beta = \infty$).
    Even though the correlation length $\xi$ of correlated environmental excitations are different by an order of magnitude, the convergence behaviour of the dynamics with respect to $D$ is the same.}
    \label{fig:CompRLR}
\end{figure}

These observations lead us to consider another hypothesis related to the connectivity of the system-environment TNS, \textit{i.e.} the number of bond dimension indices for the system in the MPS/MPO.
In the Left-Right case the system is connected twice to the environment, doubling -- so to speak -- its surface of exchange with it, whereas the Successive and Interleaved ones are coupled only once to a composite environment.
A second piece of evidence that the correlation length is not a valid explanation for the efficiency of the computation is given by a fourth mode arrangement which has a constant correlation length and two points of connection -- `interfaces' -- between the system and the environment.
Keeping a Left-Right arrangement of the mode but taking one of the chains in the reverse order (see Fig.~\ref{fig:CompRLR}~(a)) gives a situation where the correlation length is fixed (and similar to the Successive arrangement) as the excitations are now propagating in the same direction.
Hence, this new Reverse Left-Right arrangement has the same topology -- in terms of TNS connectivity -- as the Left-Right ordering but a different correlation length.
Figure~\ref{fig:CompRLR}~(b) shows the second site population for the same choice of parameters as in Fig.~\ref{fig:comparison}.
The two versions of the Left-Right arrangement have a similar convergence behaviour despite having radically different correlation lengths.
It is usually considered that long-ranged interactions are not well-handled by MPS.
Here, the fact that only one subsystem interacts with long-range couplings (i.e. the system with the environments) makes the problem tractable, further highlighting the capabilities of MPS for open systems (and more specifically in this context the TEDOPA method).

From this comparative analysis, we can rule out the hypothesis that the important metric for the MPS bond dimension is the correlation length between excitations in both environments.

Now that the hypothesis of the connection between the correlation length and the bond dimension has been refuted, it is important to notice that the intuitive relation between the dynamics of correlated excitations and the growth of bond dimension -- often invoked to justify the feasibility of a given TNS representation~\cite{rams2020breaking, kohn_quench_2022} -- is not rigorous.
As convincing as it may sound initially, there is no firm ground on which this assertion is set. 
Indeed, one could argue that dynamically correlated excitations carry very low entanglement in the MPS as the only information they share is their distance from the system which is updated by the action of the time evolution operator.

\section{Three environments: Connectivity\label{sec:3baths}}

To test the hypothesis of the importance of connectivity on the growth of bond dimensions, we study a spin interacting with three identical bosonic baths.
We name these baths `up', `left' and `right'.
The Hamiltonian of this three baths Spin Boson Model (S3BM) is
\begin{align}
\h &= \frac{\epsilon}{2}\hat{\sigma}_z + \frac{\Delta}{2}\hat{\sigma}_x + \sum_{i=1}^{3}\hat{\sigma}_x\int_{0}^{\infty}\d\omega \sqrt{J(\omega)}({\a_\omega}^i + {\a_\omega}^{i\dagger})\ ,
\end{align}
where $\hat{\sigma}_j$ are Pauli matrices, $J(\omega)$ is the bath spectral density and 
${\a_\omega}^{i\dagger}$ 
creates a bosonic excitation of energy $\omega$ in the $i$\textsuperscript{th} bath.
We keep an Ohmic spectral density with a hard cut-off $J(\omega) = 2\alpha\omega_c H(\omega_c - \omega)$.
Again, this simple model can be mapped to a SBM with a single environment, and is used here as a first investigation of our hypothesis.

\begin{figure*}
    \centering
    \includegraphics[width=\textwidth]{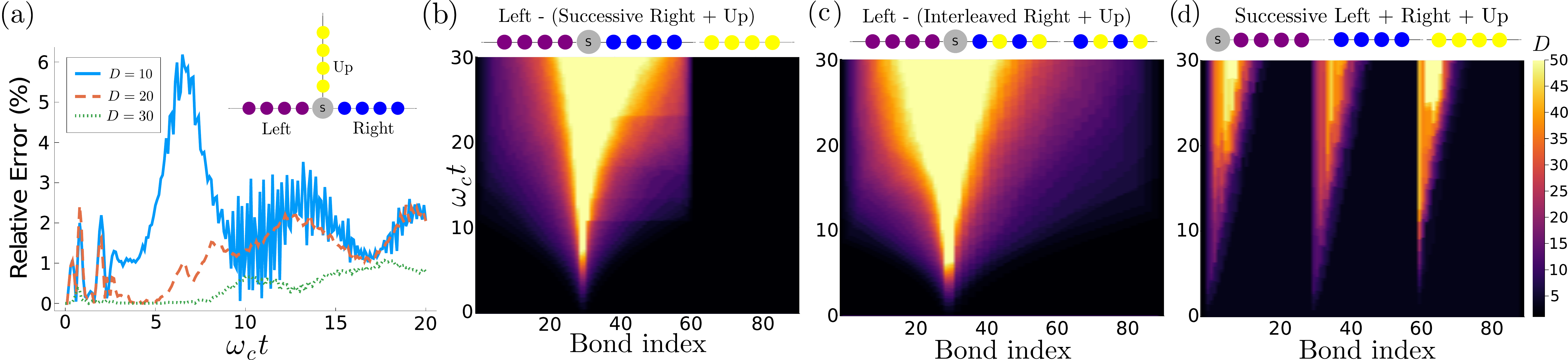}
    \caption{Impact of TNS connectivity for a TLS system interacting with three bosonic baths (S3BM) (a) Relative error between the dynamics of $\langle\sigma_z\rangle$ for different values of the bond dimension $D$ and $D_\text{Max} = 50$ obtained with 1TDVP with a TTN state. Errors of the order of a percent manifest at an early time. The simulation parameters are $\alpha = 1$, $\epsilon = 0.2\omega_c$, $\Delta = 0.5\omega_c$ and $\beta=\infty$. (b) - (d) Evolution of the bond dimensions of the MPS with: (b) a Left - S - (Successive Right + Up) configuration obtained with 2TDVP. The cut-off value for the bond dimensions $D_\text{Max} = 50$ is quickly reached and 2TDVP is not able to handle the long range coupling induced by appending the right and up chains. (c) a Left - S - (Interleaved Right + Up) configuration obtained with 2TDVP. The cut-off value for the bond dimensions $D_\text{Max} = 50$ is quickly reached. (d) a S - (Successive Left + Right + Up) configuration obtained with DTDVP. The bond dimensions evolve more slowly.}
    \label{fig:3sbm}
    \label{fig:3sbmtree}
\end{figure*}

We consider different configurations with three different system connectivity: (1) the tree configuration where the system is connected directly to three baths (see Fig.~\ref{fig:3sbm}(a)); (2) the left-right configuration where the system is connected to the left and (right + up) chains (see Fig.~\ref{fig:3sbm}(b)\&(c)); and (3) the successive configuration where the system is connected to a single chain (left + right + up) (see Fig.~\ref{fig:3sbm}(d)).
In order to study in real time the evolution of the required bond dimension necessary for the convergence of the MPS state, we time-evolve the state using a 2-site variant of the TDVP method (2TDVP)~\cite{paeckel} and the adaptive variant of the 1-site TDVP (DTDVP)~\cite{dunnett_efficient_2021}.
These two variants are able to update the bond dimensions of the MPS state in real time.
Figure \ref{fig:3sbm}~(b) shows the evolution of the bond dimension of the MPS with a Left-(Successive Right+Up) arrangement during a time evolution performed with 2TDVP.
The chains all have $N_m = 30$ modes.
We can see that the bond dimension grows quickly and reaches the cut-off bond dimension of the simulation ($D_\text{Max} = 50$) at $\omega_c t \approx 6$.
The other noticeable element is that the third chain which was appended to the right chain was not updated by 2TDVP.
This is consistent with the known fact that 2TDVP has difficulties taking into account long-ranged interactions~\cite{yang2020time}.
Hence, this case has also been studied with an Interleaved arrangement of the (right + up) chain to make it tractable with 2TDVP as shown in Fig.~\ref{fig:3sbm}~(c).
The main conclusion on the growth of the bond dimensions is unchanged.
These heat maps directly show that the growth of the bond dimensions is localised around the system.
The DTDVP algorithm which enables a one-site update of the MPS and dynamically evolving bond dimensions was also tried on this arrangement but always got stuck in the initial manifold~\cite{yang2020time}.

However, the DTDVP method was able to be used to study the one chain (Left + Right + Up) case shown in Fig.~\ref{fig:3sbm}~(d).
This chain was in a Successive arrangement and the long-range coupling was handled correctly by the method.
Here, the bond dimensions grow at a slower pace and reach maximal allowed bond dimension $D_\text{Max} = 50$ at $\omega_c t \approx 25$, which corresponds to a fourfold improvement on the case Left-(Right+Up) geometry that has a connectivity of 2.
The growth of the bond dimension with this arrangement is not localised around the system {\it per se} but at the beginning of each chain, i.e. at the point of contact between the system and the environments.

Unfortunately, the 2TDVP and DTDVP algorithms used in this work are not straightforward to apply to TTN.
In that case we went back to the 1TDVP method and chose the relative error of the expectation value $\langle \sigma_z \rangle$ with respect to the $D = 50$ results to be our metric for the growth of the bond dimension as we are not interested in the dynamics of this observable in itself.
Figure~\ref{fig:3sbmtree}~(a) shows the relative error for $D \in \{10, 20, 30\}$.
In most cases a residual error still persists in the $D=30$ case whose value is on the order of a percent.
Most importantly, these errors consistently manifest at early times $\omega_c t \lesssim 5$.
This observation points in the same direction as the results obtained for the Left-Right and Successive arrangements.

With these examples we can see that connectivity seems to play an important role in limiting the growth of the bond dimensions needed to accurately describe the state in time.
An angle of inquiry to investigate this importance of the connectivity is to reason in terms of entanglement entropy.
The entanglement entropy gives a lower bound on the required bond dimension $S = -\tr\left[\densmat\ln\left(\densmat\right)\right] \leq \ln(D)$.
When looking at a bi-partition of a quantum state, the maximal entanglement entropy $S_\text{Max}$ is set by the dimensionality $d$ of the `smallest' Hilbert space of the two reduced states on each side of the partition $S_\text{Max} = \ln\left(d\right)$.
Hence, the larger $S_\text{Max}$, the more room there is for the bond dimension to grow.
Let us call $d_S$ the local Hilbert space dimension of the system, and $d_{E}$ the local Hilbert space dimension of each of the $N_m$ environmental modes of each chain.
The dimensionality $d$ to consider for a bond between the system and the environmental modes placed to the system's right side is (i) $d = (d_E)^{N_m}$ for a tree, (ii) $d = d_S (d_E)^{N_m}$ for a Left - (Right + Up) ordering, (iii) $d = d_S$ for a fully successive or interleaved ordering.
Thus, when there is only a single `interface' between the system and its environment(s) there is less room for the bond dimension to grow.
Furthermore, one can notice that the system can only be singled out from the environments by a single bi-partition in topologies where the system is at one end of the tensor network.
However, this entanglement entropy argument should not be considered to be more than just `food for thought' as is does not explain the core of the observations presented in Fig.~\ref{fig:3sbm}, namely the dynamics of the growth of the bond dimension.

\section{Discussion\label{sec:discussion}}
The TNS geometry that reflects the underlying structure of the Hamiltonian is `intuitive' and, most of the time, it is chosen without being motivated further~\cite{de_vega_thermofield-based_2015, dunnett2021matrix, rams2020breaking}.
In this paper, we have shown that this geometry is not the most efficient one for performing OQS simulations.
When a system is interacting with several environments, different arrangements of the environmental modes -- and crucially, not only the `intuitive' TTN structures -- can lead to well-converged results.
Moreover, for the cases we have explored here via numerical experiments, counter-intuitive but simple to implement arrangements (like the Successive one where environments are concatenated) require a lower bond dimension to reach convergence.

This first result has a consequence of practical importance as it implies that joint system-environment\emph{s} state can always be written as an MPS which is easier to implement than a TTN.
The argument often given against arrangements where independent environments are not coupled directly to the system, namely that the growth of bond dimension in the multi-environments MPS state is related to the correlation length $\xi(t)$ of dynamically correlated environmental excitations, has been disproved for bosonic environments.
Showing that the usual Left-Right ordering of bath modes is less efficient than the less `natural' Successive and Interleaved arrangements leads us to consider an alternative hypothesis.
For bosonic environments, the number of bonds connecting the system to the environments (`interfaces') is a quantity associated with the growth of the bond dimension.
The lower the number of interfaces, the lower the required bond dimension.
For simple models that can be mapped easily to single environment problems, our results agree with this hypothesis.

These results have a compounding impact on the efficiency of TN simulations as the number of elements in the system's tensor scales proportionally to $D^N$ with $N$ the number of interfaces between the system and its environments.
For instance, in the S3BM the scaling is proportional to $D^3$ for TTN states, $D^2$ for Left-Right MPS, and only $D$ for Successive and Interleaved MPS.
This bond dimension $D$ will be different in these different topologies and it would be of interest to understand how these bond dimensions scale with the size of the environment.
So far, we only established a bound based on the maximum entanglement entropy and realized that this bound is minimized for topologies where a bi-partition can single out the system from the environments.
In this paper this was realized in the case of the Successive Left + Right + Up arrangement of Fig.~\ref{fig:3sbm}(d).
We note that tree topologies where the system is the head-node and connects to a single unphysical tensor that is itself connected to multiple environments would share this property.
In the future a comparison of the growth of the bond dimensions of such a tree tensor network with the Successive arrangement presented here would be of interest.
The results presented here thus imply that simple TTN states can always be recast as Successive-ordered MPS, thus {\it a priori} removing the need for structural renormalisation of environmental couplings.
These results add to the diverse body of knowledge around the importance of selecting the appropriate geometry when using TNS and highlight that the appropriate geometries are not always the ones reproducing the structure of the interactions inside the Hamiltonian.

\acknowledgments
TL, AWC and BWL thank the Defence Science and Technology Laboratory (Dstl) and Direction G\'en\'erale de l’Armement (DGA) for support through the Anglo-French PhD scheme. 
BWL acknowledges support from EPSRC grant EP/T014032/1. AWC acknowledges support from ANR project ANR-19-CE24-0028. 
All numerical results were produced using our open source Julia software package \texttt{MPSDynamics.jl}, which is freely available at \href{https://github.com/shareloqs/MPSDynamics.jl}{https://github.com/shareloqs/MPSDynamics.jl}~\cite{dunnett_angusdunnettmpsdynamics_2021, mpsdynamicsjl_2024}.

%

\end{document}